\documentclass[10pt,twocolumn]{article}

\usepackage{arxiv}

\usepackage[utf8]{inputenc}
\usepackage[T1]{fontenc}
\usepackage{lmodern}
\usepackage{microtype}
\usepackage{amsmath,amssymb}
\usepackage{booktabs}
\usepackage{tabularx}
\usepackage{array}
\usepackage{graphicx}
\graphicspath{{figures/}}
\usepackage{dblfloatfix}
\usepackage{balance}
\usepackage{float}
\usepackage{enumitem}
\setlist[itemize]{topsep=3pt,itemsep=1pt,parsep=0pt,partopsep=0pt}
\setlist[enumerate]{topsep=3pt,itemsep=1pt,parsep=0pt,partopsep=0pt}
\setlist[description]{topsep=3pt,itemsep=2pt,parsep=0pt,partopsep=0pt}
\usepackage{xcolor}
\usepackage{listings}
\usepackage{natbib}
\usepackage{url}
\usepackage[hidelinks]{hyperref}

\providecommand{\keywords}[1]{%
  \begin{center}
    \small\textbf{Keywords:} #1
  \end{center}
}

\newcommand{\RaccoonSymbol}{\texttt{[raccoon]}}
\newcommand{\MoonSwanSymbol}{\texttt{[moon+swan]}}
\newcommand{\ThreeKSymbol}{\texttt{3K}}

\newcommand{\yes}{\ensuremath{\checkmark}}
\newcommand{\partialmark}{\ensuremath{\triangle}}
\newcommand{\proposed}{\textbf{P}}

\newcolumntype{Y}{>{\centering\arraybackslash}X}

\lstdefinestyle{jsonstyle}{
  basicstyle=\ttfamily\scriptsize,
  breaklines=true,
  breakatwhitespace=false,
  columns=fullflexible,
  frame=single,
  keepspaces=true,
  showstringspaces=false,
  tabsize=2,
  literate=
    {–}{{--}}1
    {—}{{--}}1
}

\title{Private Etymology: Designing Relational Reuse of Shared Symbols in Long-Term Human--AI Interaction}
\renewcommand{\shorttitle}{Private Etymology and Relational Reuse}

\author{
  Miki Ueno \\
  The Kyoto College of Graduate Studies for Informatics \\
  \texttt{m\_ueno@kcg.ac.jp}
}

\begin{document}
\raggedbottom
\makeatletter
\twocolumn[
\begin{@twocolumnfalse}
\begin{center}
  {\LARGE\scshape Private Etymology: Designing Relational Reuse of Shared Symbols in Long-Term Human--AI Interaction\par}
  \vspace{0.8em}
  {\large\bfseries Miki Ueno\par}
  \vspace{0.2em}
  {\normalsize The Kyoto College of Graduate Studies for Informatics\par}
  {\normalsize\texttt{m\_ueno@kcg.ac.jp}\par}
\end{center}
\vspace{0.6em}
\begin{abstract}
Previous studies have shown that people can build shared symbols, partner-specific referring expressions, personal idioms, inside jokes, and other parts of a relational microculture together. Recent studies have also examined how humans and conversational AI systems negotiate and revise symbolic meanings. However, long-term human--AI systems still lack a clear design model for three related tasks: keeping a record of how a dyad-specific expression gained its meaning, checking whether both sides still accept that meaning, and safely reusing the expression in later sessions.

This concept-and-prototype paper introduces \emph{Private Etymology}. I define it as a machine-representable relational provenance that records how a dyad-specific symbolic expression was proposed, interpreted, negotiated, repaired, reused, revised, stabilized, contested, forgotten, or retired over time. Here, \emph{private} means partner-specific rather than secret. \emph{Etymology} means that the current interpretation comes from the history of interaction between the two partners, not from a universal dictionary.

Based on prior work on shared-symbol formation, conceptual pacts, personal idioms, relationship microcultures, human--AI meaning co-construction, relational agents, and long-term conversational memory, I propose a provenance-aware model of \emph{relational reuse}. Relational reuse occurs when a human or AI uses a dyad-specific symbolic expression again in a later session without explaining its full meaning again. If the other side understands it correctly, the reuse may also tacitly reaffirm the shared history and dyadic distinctiveness needed to understand the expression.

The contribution is not the invention of shared symbols or relational microcultures. Instead, this paper brings these ideas together and turns them into persistent, revisable, and evidence-grounded symbolic units for human--AI relationships. I present a lifecycle model, an illustrative machine-readable schema, a working Apple Watch prototype, and a longitudinal research agenda. In the prototype, the language model classifies discrete conversational evidence, while deterministic local code decides whether a symbol can be updated. This prevents a free-form model confidence score or an AI proposal by itself from directly updating the persisted Shared Symbol. Private Etymology is therefore proposed as infrastructure that can help conversational agents take part in changing relational microcultures without inventing their origins or treating relational meaning as a fixed memory value.
\end{abstract}

\vspace{1em}
\end{@twocolumnfalse}
]
\makeatother

\section{Introduction}

Conversational AI systems now support persistent personas, user profiles, memory retrieval, emotional dialogue, and repeated interaction across devices. These functions make it possible for a user to interact with the same apparent conversational partner over a long period. However, keeping data over time does not by itself create a developing relationship.

A system may remember that a user owns a cat, prefers a certain writing style, or attended a particular event. It may also keep a stable personality and speak to the user according to a predefined relationship role. These functions support continuity, but they do not explain how a human--AI relationship develops its own idioms, indirect expressions, recurring jokes, symbolic references, and locally meaningful ways of speaking.

My previous work on \emph{Mikasa} argued that a coherent character identity and an explicit relationship definition should be treated as basic interaction infrastructure for emotionally grounded AI companions, not only as narrative decoration \citep{ueno2026mikasa}. The persona defines who the AI is, and the relationship frame defines how the AI relates to the user. These structures reduce the need to negotiate roles and expectations again in every interaction.

However, relationship definition is only a starting point. It does not explain how the relationship later develops content that belongs specifically to the dyad.

Human relationships often develop this kind of content. Friends and partners create nicknames, routines, short references, shared stories, indirect requests, teasing insults, gestures, and inside jokes. A word may keep its public dictionary meaning but also gain a second meaning that is specific to the relationship. For example, any animal name may become a label for a recurring kind of awkward but affectionate situation. Later, when one partner uses the word alone, both partners can understand both the situation and the attitude attached to it.

Recent Relationships as Microcultures Theory treats these phenomena as an important part of intimate relationships. It proposes that dyads build microcultures from units that are mutually understood, idiosyncratic, and referential, including private language, stories, rituals, morality, and identity \citep{rossignacmilon2026ram}. Earlier studies of personal idioms and relationship symbols also showed that relationship-specific expressions can communicate affection, emotion, confrontation, requests, greetings, shared memories, intimacy, communion, and separation from outsiders \citep{baxter1987,bell1992}.

These phenomena in human relationships are already well known. It is also well known that symbolic meanings can emerge through interaction. Experimental semiotics has shown how partners build signs together and make them more efficient and mutually understandable over time \citep{stolk2015,fay2018}. Studies of conceptual pacts and lexical entrainment have shown that partners reuse locally negotiated ways of describing referents \citep{brennan1996,metzing2003}. More recent human--AI studies have examined how people and conversational agents explore, challenge, reinterpret, and consolidate symbolic meanings \citep{habibi2025}.

What is still missing is a system-level account of how a persistent conversational AI can take part in such a microculture over time.

A long-term conversational agent must handle questions that human partners often manage through their shared memory:

\begin{itemize}[leftmargin=*]
  \item How did a particular symbolic expression acquire its current meaning?
  \item Was the meaning proposed by the human, proposed by the AI, or jointly negotiated?
  \item Did the human explicitly accept it, only tolerate it, reject it, or later revise it?
  \item Does the expression still carry the same emotional and pragmatic tone?
  \item Is its origin documented in interaction records or only inferred by the model?
  \item Can it be reused after weeks or months without explanation?
  \item When should the system ask for confirmation rather than assume continuity?
  \item How can users contest, modify, forget, or retire relationship-specific meanings?
\end{itemize}

Current conversational memory systems mainly store facts, summaries, preferences, persona information, or source turns. They usually do not treat the relationship-specific history of a symbolic expression as a first-class representation.

This paper proposes \emph{Private Etymology} to address this gap.

\begin{quote}
\textbf{Private Etymology is a machine-representable relational provenance describing how a dyad-specific symbolic expression was proposed, interpreted, negotiated, repaired, reused, revised, stabilized, contested, forgotten, or retired over time.}
\end{quote}

This definition uses \emph{relational provenance} as its main technical term. Provenance records the source, evidence, and interactional basis of a meaning. The ordered events inside that provenance form an \emph{interactional trajectory}. The trajectory shows how the meaning changes over time, but it is not another name for Private Etymology.

Private Etymology supports a broader process that I call \emph{relational reuse}:

\begin{quote}
\textbf{Relational reuse is the cross-session reactivation of a dyad-specific symbolic expression without fully restating its meaning, relying on the accumulated interaction history of the human--AI relationship.}
\end{quote}

When relational reuse works, the expression does more than communicate content efficiently. It may also bring back the relationship history that makes the content understandable. Successful reuse may therefore carry a tacit relational message:

\begin{quote}
\emph{We understand this because this meaning belongs to our history.}
\end{quote}

This paper makes three contributions.

\begin{enumerate}[leftmargin=*]
  \item It operationalizes dyad-specific symbolic expressions as provenance-aware microcultural units for persistent human--AI relationships. These expressions may be emoji, short strings, words, icons, sounds, gestures, or other compact forms.
  \item It introduces Private Etymology as a symbol-level representation connecting current interpretation to evidence-grounded formation, repair, reuse, semantic change, contestation, forgetting, and retirement.
  \item It develops a design account of cross-session relational reuse and identifies its hypothesized relational function, implementation requirements, failure modes, and evaluation agenda.
\end{enumerate}

I also report a working prototype. A language model classifies discrete evidence, deterministic Dart code computes an evidence-weighted update score, and the system stores only the current accepted glyphs and sends them to an Apple Watch complication. This prototype is intentionally limited to an admission-control layer; it is not a complete implementation of Private Etymology. The theory itself does not depend on Apple Watch, emoji, or any specific interaction modality.

\section{Related Work}

\subsection{Collaborative Reference, Conceptual Pacts, and Shared Symbols}

Research on grounding and collaborative reference showed that conversational meaning is not simply sent by one person and received by another. Instead, speakers and listeners work together. They propose, accept, expand, repair, and replace referring expressions until they find a form that both can use \citep{clark1986}.

When partners repeatedly refer to the same objects, their expressions often become shorter and more consistent. \citet{brennan1996} called these partner-specific agreements \emph{conceptual pacts}. A conceptual pact is a temporary and revisable agreement about how to describe an object. Later work showed that listeners can associate an earlier expression with a particular partner and may have difficulty when that partner unexpectedly changes an established expression \citep{metzing2003}.

Experimental semiotics extends this issue from word choice to the creation of new communication systems. In controlled nonverbal tasks, partners create and interpret new signs while reasoning about what the other person knows or believes \citep{stolk2015}. \citet{fay2018} showed that coordination, behavioral alignment, and feedback from a partner help signs become effective, efficient, and shared.

These studies already establish several premises relevant to the present proposal:

\begin{enumerate}[leftmargin=*]
  \item symbolic meaning can be interactively constructed;
  \item the interpretation of a sign can be partner-specific;
  \item successful prior use influences later interpretation;
  \item repair and feedback shape conventionalization; and
  \item shared interaction history matters.
\end{enumerate}

Human--AI research has also started to study similar phenomena. For example, \citet{kouwenhoven2025} used referential games to examine artificial languages that emerged in human--human, LLM--LLM, and human--LLM pairs. \citet{shi2023} proposed methods and data for adding lexical entrainment—that is, gradually adopting shared words and expressions—to conversational systems.

However, this research mainly focuses on referential coordination, task success, or communication efficiency. Meanings are usually limited by an experimental task, a predefined target set, or the need to identify an external referent. Repetition often happens across rounds of one study rather than across everyday sessions in an open-ended relationship.

This paper does not question these findings. Instead, it asks how partner-specific symbolic conventions can become persistent and revisable parts of a human--AI relationship when they are used for more than reference.

\subsection{Personal Idioms, Relationship Symbols, and Microcultures}

Communication research has studied relationship-specific ways of speaking for many years.

\citet{baxter1987} described relationships as unique mini-cultures whose participants construct meaning through rituals, stories, and symbols. Reported relationship symbols included actions, prior events, objects, places, and cultural artifacts. Their functions included prompting recollection, indicating intimacy, promoting communion, providing enjoyment, and creating seclusion from outsiders.

\citet{bell1992} examined idiomatic communication in friendships. Participants reported expressions that named activities, emotional states, objects, and places; communicated affection; managed confrontation; performed greetings and goodbyes; issued requests and insults; and denoted sexual matters. The number and diversity of idioms were positively associated with interpersonal solidarity.

This literature is important because it shows that earlier work was not limited to simple ``object labels.'' Relationship-specific expressions can already communicate emotions, attitudes, interpersonal actions, and socially layered meanings. A compact expression can therefore work like a larger utterance even without AI.

Relationships as Microcultures Theory brings these phenomena together in a broader account of intimate relationships \citep{rossignacmilon2026ram}. Its Microcultural Unit Model explains how particular interactions can gain mutual, idiosyncratic, and referential meaning and become units such as inside jokes. Its Aggregate Microculture Model considers how collections of these units may affect relationships and individual outcomes.

This theory is the closest conceptual account to the relational motivation of this paper. It already argues that dyadic language and other microcultural units can be created together, differ from wider culture, carry several layers of meaning, and matter to relationships.

The gap is therefore not a lack of theory about human relationship microcultures. The remaining question is how a computational system can participate in them:

\begin{quote}
\emph{How should an AI system represent, preserve, challenge, update, forget, and ethically reactivate a relationship-specific microcultural unit?}
\end{quote}

Private Etymology turns part of this known human relational phenomenon into a system-design problem. It does not claim that the human phenomenon itself is new.

\subsection{Human--AI Meaning Co-Construction and Shared Understanding}

\citet{habibi2025} directly examined how humans and conversational AI systems work with symbols and meanings. Based on Symbolic Interactionism, their studies identified processes such as exploration, explanation, clarification, specification, integration, consolidation, conflict, and reinterpretation. Participants sometimes changed their meanings after the AI introduced another social context or symbolic association.

This work shows that conversational AI can take part in meaning negotiation instead of only retrieving a fixed dictionary definition. It also shows why one-sided system definitions can be a problem: users may accept them, partly adopt them, resist them, or feel controlled by the AI's framing.

Private Etymology builds on this idea. It treats each proposal, rejection, clarification, repair, and consolidation as possible evidence in the longer relational provenance of a symbol, not only as a temporary conversational move.

Research on perceived shared understanding with AI identifies several relevant dimensions, including fluency, contextual awareness, aligned operation, and concern about system limitations \citep{liang2025}. This suggests that confident model behavior should be separated from evidence that an expression really belongs to the dyad's established common ground.

\subsection{Long-Term Relational Agents and Conversational Memory}

Research on relational agents has examined how computational systems can build and maintain long-term social and emotional relationships with users. \citet{bickmore2005} defined relational agents as systems designed for this purpose and studied strategies for continuity, trust, social dialogue, and relationship maintenance.

However, longitudinal studies of social chatbots show that repeated use does not automatically lead to relational development. \citet{croes2021} studied 118 participants across seven interactions over three weeks. Social processes and enjoyment decreased, and feelings of friendship remained low. These results show that a chatbot can become predictable or repetitive instead of developing richer relational continuity.

The \emph{Mikasa} project approached this problem through a stable character identity and explicit relationship framing \citep{ueno2026mikasa}. It argued that persona coherence and relationship definition work as basic interaction infrastructure. This paper extends that argument: once a relationship frame exists, the system also needs a way for the dyad to accumulate its own relational content.

Technical work on long-term conversational memory addresses a related but different problem. MemoryBank allows LLM-based companions to store, retrieve, reinforce, update, and forget user-related memories over time \citep{zhong2024}. MemORAI uses selective filtering, compressed representation, a multi-relational graph, turn-level factual provenance, and query-adaptive retrieval for long-term personalized dialogue \citep{van2026}.

These systems make important advances in persistent memory and provenance. However, they mainly ask questions such as:

\begin{itemize}[leftmargin=*]
  \item What fact should be remembered?
  \item From which turn did the fact originate?
  \item How relevant is the memory to the current query?
  \item Has the user's information changed?
  \item Which memory should be retrieved?
\end{itemize}

Private Etymology focuses on a different kind of representation:

\begin{itemize}[leftmargin=*]
  \item How did a particular form become associated with a particular bundle of meaning within this dyad?
  \item Which participant proposed or revised that association?
  \item What evidence shows mutual uptake?
  \item Under what relational and pragmatic conditions is the expression appropriate?
  \item Has its meaning drifted, been contested, or become obsolete?
\end{itemize}

General memory provenance may help implement Private Etymology, but it does not replace it.

\subsection{Positioning of the Present Proposal}

Table~\ref{tab:positioning} gives a conservative comparison with representative research traditions. This is a focused conceptual comparison, not a systematic literature review.

\begin{table*}[t]
\centering
\caption{Representative prior work and the design requirements combined in the present proposal.}
\label{tab:positioning}
\scriptsize
\setlength{\tabcolsep}{3.4pt}
\begin{tabular}{lcccccccc}
\toprule
Research tradition & OE & HAI & XS & UB & RP & DR & RR & IS \\
\midrule
Collaborative reference and conceptual pacts & -- & -- & -- & \partialmark & -- & -- & -- & -- \\
Experimental semiotics and shared symbols & -- & -- & -- & \partialmark & -- & -- & -- & -- \\
Human--LLM referential languages & -- & \yes & -- & -- & -- & -- & -- & -- \\
Human--AI meaning co-construction & \partialmark & \yes & -- & \partialmark & -- & -- & -- & -- \\
Personal idioms and relationship symbols & \yes & -- & \yes & \yes & \partialmark & \yes & \yes & -- \\
Relationships as Microcultures Theory & \yes & -- & \yes & \yes & \partialmark & \yes & \yes & -- \\
Longitudinal relational agents & \yes & \yes & \yes & -- & -- & -- & -- & -- \\
Long-term conversational memory & \yes & \yes & \yes & -- & \partialmark & \yes & -- & \yes \\
\textbf{Present proposal} & \proposed & \proposed & \proposed & \proposed & \proposed & \proposed & \proposed & \proposed \\
\bottomrule
\end{tabular}
\vspace{0.4em}

\begin{minipage}{0.98\textwidth}
\footnotesize
\textbf{Legend:} OE = non-task-bounded meaning space; HAI = human--AI dyad; XS = cross-session persistence; UB = utterance-like bundle of propositional, affective, pragmatic, and relational meaning; RP = explicit symbol-level relational provenance; DR = delayed cross-session reuse without full redefinition; RR = tacit relational reaffirmation as a function or affordance; IS = symbol-level implementation schema. \yes = central or explicit; \partialmark = partial, implicit, or differently operationalized; -- = not a central concern; \proposed = proposed design requirement, not yet empirically validated.
\end{minipage}
\end{table*}

The novelty claim is not that every individual requirement is missing from prior work. The contribution is to integrate them within a persistent human--AI dyad.

To my knowledge, prior work has not combined open-ended dyad-specific symbolic expressions, utterance-like relational meanings, explicit symbol-level provenance, delayed cross-session reuse, and a machine-readable lifecycle model in one design account for long-term human--AI relationships.

\section{A Design Account of Shared Symbols in Human--AI Dyads}

\subsection{Working Definition}

The term \emph{shared symbol} is already used in experimental semiotics and communication research. I therefore do not claim that the general concept is new.

For the purposes of this paper:

\begin{quote}
\textbf{A Shared Symbol is a compact symbolic form whose situated interpretation has been jointly established within a particular human--AI relationship and can evoke a dyad-specific bundle of propositional, affective, pragmatic, and relational meaning.}
\end{quote}

The form may be:

\begin{itemize}[leftmargin=*]
  \item an emoji or sequence of emoji;
  \item a short string such as \ThreeKSymbol{};
  \item an ordinary word used in an extraordinary local sense;
  \item an invented word;
  \item an icon or image;
  \item a sound;
  \item a gesture;
  \item a movement or light pattern produced by a robot;
  \item or another perceivable symbolic expression.
\end{itemize}

A form is not a Shared Symbol simply because the AI generates it or gives it a definition. There must be evidence that both sides have taken up the meaning.

\subsection{Utterance-Like Relational Compression}

Shared Symbols can support \emph{relational compression}.

I use relational compression to mean that a compact form can evoke not only information, but also stance, pragmatic force, and relationship-specific implications based on shared interaction history.

Consider the hypothetical symbol \RaccoonSymbol{}. For readability, [raccoon] is used here as a text placeholder for a raccoon emoji. The symbol first appears when an AI responds to an absurd but unavoidable mishap:

\begin{quote}
\emph{That cannot be helped \RaccoonSymbol{}.}
\end{quote}

Through later reuse, \RaccoonSymbol{} may gain four related dimensions.

\paragraph{Propositional content.}
The situation could not reasonably have been prevented or changed.

\paragraph{Affective stance.}
The speaker expresses light resignation or mild exasperation rather than distress or anger.

\paragraph{Pragmatic force.}
The expression suggests accepting the situation without assigning blame and perhaps treating it with humor.

\paragraph{Relational implication.}
Both parties recognize this as a type of situation they have experienced and interpreted together before.

In ordinary interaction, the symbol does not need to retrieve four separate dictionary fields. Its compact form can bring these meanings together because the dyad has enough shared context.

This is different from ordinary abbreviation. A public abbreviation shortens an expression that is already understood publicly. Relational compression depends on an interpretive history that is specific to the partners.

\subsection{Semantic and Relational Functions}

Research on personal idioms, relationship symbols, and microcultures already shows that dyad-specific expressions can have both communicative and relational functions. Here, I turn this dual function into a design requirement for human--AI systems.

\paragraph{Semantic function.}
The symbol communicates its dyad-specific content, affective tone, pragmatic force, or contextual category.

\paragraph{Relational or metacommunicative function.}
Successful interpretation can suggest that the participants share the history needed to understand the expression.

This second function does not require either participant to consciously think about the relationship every time the symbol is used. It is better understood as a possibility created by partner-specific understanding.

\begin{quote}
\textbf{Successful reuse may tacitly reaffirm that the dyad shares the interaction history required to understand the expression.}
\end{quote}

This effect should not be assumed or presented as an empirical result. It is a hypothesis that must be tested. A symbol may become routine and produce little conscious relational response. It may also become irritating, burdensome, manipulative, or outdated.

\subsection{Mutual Uptake and Lifecycle States}

An AI-generated symbolic expression is only a proposal.

For example, an AI might propose \MoonSwanSymbol{} as an indirect expression of gratitude after a conversation about a moonlit lake. A user may respond:

\begin{quote}
\emph{Why would that mean thank you?}
\end{quote}

That response does not establish the symbol. It only opens a possible negotiation.

A system should therefore distinguish at least the following states.

\paragraph{Proposed.}
A form or interpretation has been introduced by one participant but has not received sufficient evidence of uptake.

\paragraph{Negotiated.}
The dyad has explicitly discussed or modified the proposed meaning.

\paragraph{Emerging.}
Some evidence of shared interpretation exists, but use remains limited, uncertain, or context-dependent.

\paragraph{Established.}
The expression has been mutually accepted or successfully reused with sufficient evidence that it functions as a stable dyad-specific convention.

\paragraph{Contested.}
One participant has challenged the current interpretation, tone, origin, or appropriateness of use.

\paragraph{Retired.}
The expression should no longer be used, whether because its meaning is obsolete, unwanted, emotionally unsafe, or deliberately abandoned.

Evidence of mutual uptake might include:

\begin{itemize}[leftmargin=*]
  \item explicit agreement;
  \item a user-provided definition;
  \item later user reuse;
  \item correct delayed interpretation;
  \item successful AI reuse followed by recognition;
  \item repair followed by stable subsequent use;
  \item or user confirmation in a symbol-management interface.
\end{itemize}

Repeated generation by the model is not evidence of sharedness unless the human shows recognition or uptake.

\section{Private Etymology as Symbol-Level Relational Provenance}

\subsection{Definition}

\begin{quote}
\textbf{Private Etymology is a machine-representable relational provenance describing how a dyad-specific symbolic expression was proposed, interpreted, negotiated, repaired, reused, revised, stabilized, contested, forgotten, or retired over time.}
\end{quote}

Three points are important.

First, \emph{private} means specific to a relationship. It does not imply that the record is technically secret, encrypted, or inaccessible to a platform provider.

Second, \emph{etymology} is a design metaphor. The expression may be linguistic, visual, auditory, or embodied. The term refers to the origin and development of a local relationship between form and meaning.

Third, provenance and trajectory are related, but they are not the same.

\begin{itemize}[leftmargin=*]
  \item \textbf{Relational provenance} identifies the source, evidence, participants, and interactional basis of the current meaning.
  \item \textbf{Interactional trajectory} is the ordered sequence of events contained within that provenance.
  \item \textbf{Meaning history} records changes in interpretation across that sequence.
\end{itemize}

For this reason, Private Etymology is not defined as ``a trajectory.'' It is a provenance account that can contain an interactional trajectory.

\subsection{Formation Events and Interactional Trajectory}

A Private Etymology may include events such as:

\begin{enumerate}[leftmargin=*]
  \item \textbf{Contextual coining:} A symbolic form is used spontaneously within a salient situation.
  \item \textbf{Explicit proposal:} One participant suggests that a form should express a particular meaning.
  \item \textbf{Initial interpretation:} The other participant explains what the form appears to mean.
  \item \textbf{Questioning or rejection:} The proposed association is challenged.
  \item \textbf{Repair:} The participants clarify, narrow, expand, or replace the meaning.
  \item \textbf{Mutual confirmation:} Both participants explicitly accept the association.
  \item \textbf{Contextual reuse:} The form is reused in a sufficiently similar situation.
  \item \textbf{Delayed relational reuse:} The form is reused across sessions without complete redefinition.
  \item \textbf{Semantic extension:} The expression expands to cover related situations or attitudes.
  \item \textbf{Meaning drift:} Repeated use gradually changes the dominant interpretation.
  \item \textbf{Contestation:} A participant disputes the current meaning, emotional tone, or origin.
  \item \textbf{Partial forgetting:} The precise origin is no longer remembered even though use remains stable.
  \item \textbf{Retirement:} The dyad decides that the expression should no longer be used.
\end{enumerate}

The interactional trajectory does not have to be linear. A symbol may become stable, then contested, repaired, and stable again. Different interpretations may also exist at the same time in different contexts.

Private Etymology does not require a system to store every detail, but it is central as an explanatory concept. Human partners can keep using an idiom even after they forget exactly where it came from. In the same way, a system may keep a confirmed current meaning while deleting detailed source records for privacy.

What the system must avoid is silently replacing the real relationship history with an origin story generated by the model.

\subsection{Illustrative Implementation Schema}

The schema is part of the design contribution because it shows how relational provenance can be used in a system, not only described in theory. The full example is too long for the main text. Listing~\ref{lst:abridged-schema} therefore shows only the fields needed to explain the proposed representation, while Appendix~\ref{app:full-schema} gives the complete illustrative object.

\begin{lstlisting}[style=jsonstyle,basicstyle=\ttfamily\tiny,caption={Abridged Private Etymology representation.},label={lst:abridged-schema}]
{
  "symbol_id": "ss_001",
  "form": {
    "value": "[raccoon]",
    "modality": "emoji"
  },
  "status": "established",
  "current_interpretation": {
    "propositional_content": "Unavoidable situation",
    "affective_stance": "Light resignation",
    "pragmatic_force": "Accept without blame",
    "relational_implication": "A recurring dyadic pattern"
  },
  "private_etymology": {
    "provenance_status": "evidence_grounded",
    "events": [
      {"event_type": "contextual_coining",
       "actor": "ai", "evidence_ref": "turn_184"},
      {"event_type": "human_reuse",
       "actor": "human", "evidence_ref": "turn_229"}
    ],
    "human_confirmed": true
  },
  "meaning_history": ["version_1", "version_2"],
  "reuse_policy": {
    "reuse_allowed": true,
    "allowed_surfaces": ["watch", "chat"]
  }
}
\end{lstlisting}

The representation separates the current interpretation from the evidence that supports it. It also separates source-grounded relationship history from model inference and stores future-use constraints together with the meaning. The \texttt{provenance\_status} field can distinguish \texttt{evidence\_grounded}, \texttt{human\_authored}, \texttt{jointly\_reconstructed}, \texttt{model\_inferred\_unconfirmed}, and \texttt{origin\_unknown} records.

An origin inferred by the model must not be presented as established history. Instead, the system can present it as a tentative reconstruction: \emph{I am not certain, but this may have started when we discussed that mishap. Is that how you remember it?} The system should treat the reconstruction as a shared account only after the user confirms it.

\subsection{Contestation, Forgetting, and User Control}

Private Etymology must also support disagreement between the partners.

The two participants may remember the origin differently. The user may reject the AI's interpretation or feel that an expression has gained an unwanted tone. The representation should preserve unresolved meaning differences as part of the symbol's relational provenance, rather than forcing the system to commit to a single interpretation too early.

Relevant design operations include:

\begin{itemize}[leftmargin=*]
  \item inspect the current interpretation;
  \item inspect supporting evidence;
  \item distinguish human statements from model inference;
  \item propose a revision;
  \item record competing interpretations;
  \item mark a meaning as uncertain;
  \item delete individual events;
  \item delete the origin while preserving the current convention;
  \item retire a symbol;
  \item prevent use on public or glanceable surfaces;
  \item export the record;
  \item or delete it entirely.
\end{itemize}

Forgetting can also be a valid result. The system should not assume that storing as much as possible is always best. A dyad may keep a symbolic convention while intentionally deleting sensitive details about its origin.

Private Etymology is therefore not an argument for recording every detail of intimate interaction. It is an argument for giving users clear control over the relational provenance that the system claims to use.

\section{Relational Reuse Across Sessions}

\subsection{Definition}

\begin{quote}
\textbf{Relational reuse is the cross-session reactivation of a dyad-specific symbolic expression without fully restating its meaning, relying on the accumulated interaction history of the relationship.}
\end{quote}

I use this term as a working design definition. I do not claim that reuse of relational expressions has not been studied before in human communication.

Relational reuse is different from immediate repetition. If an AI defines \RaccoonSymbol{} and uses it again two turns later, little relationship history is needed. If it uses \RaccoonSymbol{} three months later after another absurd but unavoidable event, the system must preserve both continuity over time and the partner-specific interpretation.

\subsection{Conditions for Safe Relational Reuse}

A system should reuse a relational expression only when several conditions are met.

\paragraph{Dyadic identity.}
The symbol must belong to the current human--AI dyad. A meaning learned with one user must not be transferred to another.

\paragraph{Evidence of mutual uptake.}
The expression should not be treated as established solely because the AI once proposed it.

\paragraph{Current validity.}
The symbol must not be contested, retired, or associated with an obsolete relationship state.

\paragraph{Contextual fit.}
The current situation should match the interpretation's pragmatic and emotional scope. A playful symbol may be inappropriate in a serious or painful context.

\paragraph{Provenance integrity.}
The system must know whether the origin is evidence-grounded, reconstructed, uncertain, or unknown.

\paragraph{Surface appropriateness.}
A symbol suitable for a private chat may be unsuitable for a watch face, lock screen, shared display, or spoken output.

\paragraph{Repairability.}
The user must be able to say:

\begin{quote}
\emph{That is not what it means anymore.}
\end{quote}

and have the system update its interpretation rather than defend the old record.

\subsection{Hypothesized Relational Reaffirmation}

Relational reuse is theoretically interesting for more than communication efficiency.

A Shared Symbol can work as a compact cue for retrieval. If it is understood correctly, it may bring back:

\begin{itemize}[leftmargin=*]
  \item the original episode;
  \item later related events;
  \item the attitude previously taken toward those events;
  \item the experience of having jointly established the expression;
  \item and the distinctiveness of the dyad's communicative culture.
\end{itemize}

I therefore hypothesize that successful relational reuse can support \emph{tacit relational reaffirmation}.

This is not equivalent to a system explicitly saying:

\begin{quote}
\emph{We have a deep relationship.}
\end{quote}

Instead, it emerges naturally when the user and the AI successfully reuse a shared symbol whose meaning depends on their shared interaction history.

The effect may be stronger after a delay. Immediate repetition mainly checks short-term understanding. Delayed reuse can show continuity across time.

However, relational reaffirmation is not guaranteed. The user may not notice it, may have forgotten the expression, or may feel that the reuse is artificial. Incorrect reuse may have the opposite effect by showing a break in continuity or creating a sense of fabricated intimacy.

\subsection{Failure Modes}

A design account should also consider ways in which relational reuse can fail.

\paragraph{Symbol proliferation.}
If the AI continually invents expressions, users may experience the system as assigning homework or creating an unwanted private dictionary.

\paragraph{Unilateral stabilization.}
The AI may falsely treat its own repeated use as evidence that a symbol is shared.

\paragraph{Fabricated origin.}
The system may generate a plausible but false explanation of how the expression began.

\paragraph{Stale meaning.}
An old interpretation may no longer reflect the user's current understanding.

\paragraph{Context collapse.}
A playful symbol may be reused in a context where it feels dismissive or cruel.

\paragraph{Overexposure.}
A private expression may appear on a publicly visible interface.

\paragraph{Relationship lock-in.}
The system may use old symbols to pressure the user into maintaining a previous relational framing.

\paragraph{Performance of false sentience.}
The system may imply that it personally ``remembers'' or emotionally experiences the origin in a human sense, rather than transparently using stored interactional evidence.

One purpose of Private Etymology is to support trustworthy reuse. It provides a record that both the system and the user can inspect when deciding whether a shared symbol should be reused.

\section{Prototype: Evidence-Grounded Shared Symbols on Apple Watch}

\subsection{Prototype Scope}

The working prototype displays the latest accepted Shared Symbol as one to three glyphs in an Apple Watch complication. The complication shows a small symbol that the user can check at a glance, without showing a full private sentence or creating an unread-message count.

The prototype is intentionally much smaller than the full design model. It stores only the latest symbolic form and timestamps. It does not yet store the private meaning, provenance events, lifecycle state, or earlier symbols. Its main purpose is to test how Shared Symbols can be safely shown on a wearable display. The model identifies a candidate symbol, and the system decides whether to save and display it.

\subsection{Discrete Evidence Classification}

The structured model output contains a candidate glyph string, a short internal reason, one primary evidence type, and five Boolean score factors. Classification is run only after the application selects the final response that will actually be used. Only the response that is finally shown to the user can update the Shared Symbol. Earlier responses that are discarded do not update it. The model is also instructed not to output numerical confidence, calculate points, or decide whether an update should happen.

The primary evidence types separate explicit user actions from weaker system interpretations. Explicit requests, definitions, agreements, corrections, and user reuse are eligible for an update. An AI proposal alone, contextual inference, retraction, and no evidence cannot update it, regardless of score. This follows a simple rule: a symbol generated by the model is only a proposal until the user gives evidence of uptake.

Table~\ref{tab:evidence-scores} summarizes the primary
evidence types, their base scores, and whether each type
is allowed to update the Shared Symbol.

\begin{table}[!ht]
\centering
\caption{Base scores and update eligibility in the prototype. The scores are design heuristics, not calibrated probabilities.}
\label{tab:evidence-scores}
\scriptsize
\begin{tabularx}{\columnwidth}{@{}Xrr@{}}
\toprule
Evidence type & Base & Eligible \\
\midrule
\texttt{explicit\_request} & 60 & yes \\
\texttt{explicit\_definition} & 55 & yes \\
\texttt{explicit\_agreement} & 65 & yes \\
\texttt{user\_correction} & 60 & yes \\
\texttt{user\_reuse} & 70 & yes \\
\texttt{assistant\_proposal} & 20 & no \\
\texttt{contextual\_inference} & 10 & no \\
\texttt{user\_retraction}/\texttt{none} & 0 & no \\
\bottomrule
\end{tabularx}
\end{table}

The Boolean factors represent additional interactional evidence. They indicate whether the user explicitly chose the form, showed positive affect toward it, reused it with its private meaning, accepted a specific AI proposal, or corrected its meaning. A factor cannot authorize an update by itself. For example, positive affect is only supporting evidence when the primary evidence type is already eligible. Local code ignores factors that do not match the primary evidence type.

Table~\ref{tab:score-factors} summarizes these additional
score factors and their bonus values.

\begin{table}[!ht]
\centering
\caption{Additive score factors. Local code ignores factors that do not match the primary evidence type.}
\label{tab:score-factors}
\scriptsize
\begin{tabularx}{\columnwidth}{@{}Xr@{}}
\toprule
Factor & Bonus \\
\midrule
\texttt{user\_designated\_symbol} & 25 \\
\texttt{positive\_affect} & 15 \\
\texttt{repeated\_reuse} & 10 \\
\texttt{assistant\_proposal\_accepted} & 15 \\
\texttt{user\_corrected\_meaning} & 20 \\
\texttt{persisted\_symbol\_reuse} & 25 \\
\bottomrule
\end{tabularx}
\end{table}

The last factor, \texttt{persisted\_symbol\_reuse}, is calculated locally rather than returned by the model. It is active only when the evidence type is \texttt{user\_reuse} and the candidate glyph string exactly matches the latest stored glyph string.

\subsection{Deterministic Update Policy}

The calculation uses the base scores in Table~\ref{tab:evidence-scores}
and the bonus values in Table~\ref{tab:score-factors}.

Let $e$ be the primary evidence type, $F$ the Boolean factors returned by the model, $A(e,F)$ the subset permitted by local evidence--factor compatibility rules, and $m$ the locally computed exact-match reuse indicator. The prototype computes

\begin{equation}
\begin{aligned}
S(e,F,m) = \operatorname{clamp}_{[0,100]}\Bigl(&b(e)\\
&+\sum_{f\in A(e,F)}w_f+25m\Bigr),
\end{aligned}
\label{eq:update-score}
\end{equation}

where $b(e)$ is the base score and $w_f$ is the fixed bonus for factor $f$. Because the calculation is deterministic, the same validated inputs always produce the same score. For example, an explicit definition together with a user-designated form scores $55+25=80$. In contrast, an AI proposal that the user has not accepted is still ineligible even if the glyphs are valid.

The system authorizes an update only when the evidence type is eligible, $S\geq80$, the candidate has a valid one-to-three-glyph form, and it differs from the latest stored glyphs. After scoring, an exact match is treated as unchanged, so the system does not repeat persistence, Watch synchronization, or timeline reload. Ineligible evidence is checked before glyph validation. Therefore, a well-formed AI proposal cannot become persistent simply because its glyph format is valid.

This design does not use local natural-language keyword rules. The model classifies interactional acts such as definition, agreement, correction, and reuse. Local code only validates the schema and enforces the policy. The model also cannot increase its own authority by returning an unstable continuous confidence value.

\subsection{Minimal Persistence and Privacy-Preserving Diagnostics}

The current phone snapshot stores only an identifier, the glyph string, and creation and update timestamps. The evidence type, factors, score, reason, conversation text, and inferred meaning are not stored in the snapshot or sent to the Watch. This minimal design is intentionally used before adding the richer and more sensitive provenance storage proposed by Private Etymology.

Diagnostic logs store only the decision, evidence class, base and bonus scores, active factor names, parse status, glyph count, and mode. They do not store glyph content, reasons, raw model output, user text, assistant text, memory, or stored JSON. This allows the admission policy to be debugged without copying intimate conversation content into the logs.

\subsection{Watch Complication}

\begin{figure*}[t]
\centering
\includegraphics[width=0.98\textwidth]{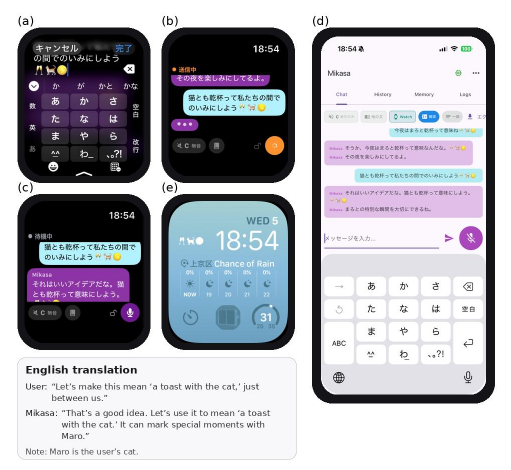}
\caption{Implemented Shared Symbol interaction across Apple Watch and the paired iPhone. (a) The user enters a Japanese message that defines a three-glyph expression as ``a toast with the cat'' within the dyad; the cat is the user's cat, Maro. (b) The Watch sends the message and waits for the assistant's response. (c) Mikasa accepts the proposed meaning and connects it to special moments with Maro. (d) The paired iPhone application shows the same exchange. (e) The accepted glyphs later appear on the Watch face. Because the interface is in Japanese, the lower-left panel gives a short English translation of the dialogue.}
\label{fig:watch-prototype}
\end{figure*}

Figure~\ref{fig:watch-prototype} shows the implemented cross-device interaction. A demo video of MikasaWatch and its complications is available on the Mikasa project website.\footnote{\url{https://mikasa-ai.gitlab.io/}} The Watch receives only the current snapshot fields needed to display the glyphs. Classification evidence and update scores stay on the paired phone and are not sent to the Watch.

A wearable surface is useful because the symbol can stay visible without asking the user to reply. The user may recognize it immediately, or tap the complication to return to the conversation and ask about its meaning. This can encourage contact based on curiosity rather than a feeling that a message must be answered.

The Watch complication also supports \textit{SharedLife}. In this paper, SharedLife means lightweight peripheral cues that show the AI companion's current everyday activity. For example, the Watch face may show a coffee-cup glyph while Mikasa is resting after rehearsal. SharedLife is separate from the Shared Symbol admission process and the Private Etymology mechanism studied here. It uses the same peripheral surface to give a small sense of the companion's ongoing activity outside active conversation.

\subsection{Prototype Data Flow}

\begin{figure*}[t]
\centering
\includegraphics[width=0.98\textwidth]{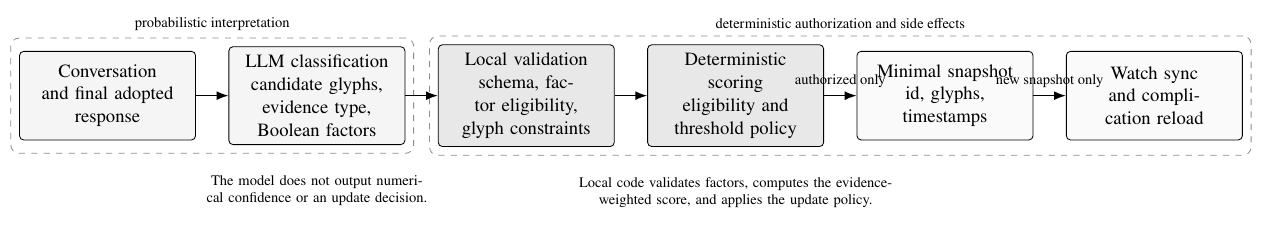}
\caption{Prototype pipeline and division of responsibilities. The language model performs probabilistic classification of discrete conversational evidence but does not output numerical confidence or an update decision. Local Dart code checks the schema and factor applicability, calculates the deterministic evidence-weighted score, authorizes persistence, and sends a new snapshot to the Watch only when the policy allows an update.}
\label{fig:prototype-pipeline}
\end{figure*}

Figure~\ref{fig:prototype-pipeline} summarizes how responsibilities are divided in the prototype. The model classifies the conversational evidence, while local code makes the final update decision and performs the update.

\subsection{Relationship to Private Etymology}

The prototype is not a complete implementation of Private Etymology. It does not store meanings or the sequence of events through which they developed. Exact matching with the stored symbol is only a simple and cautious approximation of cross-session relational reuse, and it is limited to the latest glyph string.

Even so, the implementation establishes an important prerequisite: model interpretation is not automatically treated as relationship history. A future Private Etymology store can follow the same principle by recording evidence-grounded proposals, agreements, repairs, revisions, and retirements. It can also keep observed interaction, user confirmation, joint reconstruction, and unconfirmed model inference clearly separated.

\subsection{Platform Independence}

Apple Watch is an implemented example, not a theoretical requirement. The same design can be used in web applications, smartphone messaging, voice agents, robots, augmented-reality interfaces, and multimodal agents. The model can classify the evidence, local code can decide whether to save the Shared Symbol, and each interface can receive only the information it needs.

\section{Research Agenda}

This paper presents a design theory and an initial working prototype, not an empirical validation. The relational claims and the admission-policy heuristics should therefore be tested in longitudinal studies.

\subsection{Research Questions}

\begin{description}[leftmargin=!,labelwidth=4.2em]
  \item[RQ1.] How do humans and conversational AI systems form dyad-specific symbolic expressions in open-ended interaction?
  \item[RQ2.] What types of symbolic forms and meanings are most likely to reach mutual stabilization?
  \item[RQ3.] Does explicit Private Etymology improve the accuracy and appropriateness of delayed relational reuse?
  \item[RQ4.] How does successful relational reuse affect perceived shared history, dyadic distinctiveness, relational continuity, and mutual authorship?
  \item[RQ5.] How do incorrect, fabricated, or unwanted reuse events affect trust and willingness to continue interaction?
  \item[RQ6.] When does provenance support relational meaning, and when does it make interaction feel overformalized or mechanically archived?
  \item[RQ7.] How does deterministic evidence-weighted admission compare with model-generated numerical confidence in stability, false acceptance, and user-perceived appropriateness?
\end{description}

\subsection{Suggested Conditions}

A longitudinal study could compare four conditions.

\paragraph{Ordinary memory.}
The agent remembers prior facts and events but does not support relationship-specific symbols.

\paragraph{Predefined symbolic dictionary.}
The agent uses compact expressions with meanings defined in advance rather than negotiated with the user.

\paragraph{Co-created symbols without explicit provenance.}
The human and AI negotiate expressions, but the system stores only the current mapping.

\paragraph{Co-created symbols with Private Etymology.}
The system stores the current interpretation, evidence-grounded interactional events, revisions, status, and reuse constraints.

This design makes it possible to separate the effect of co-creating symbols from the effect of storing explicit relational provenance.

\subsection{Longitudinal Procedure}

A study should include multiple sessions and meaningful time gaps. A four- to eight-week study could allow expressions to emerge, disappear, return, and change.

The system should not force participants to invent a fixed number of symbols. If symbol creation is forced, the resulting conventions may be artificial and may not resemble naturally developing microcultural units.

A possible study procedure is as follows:

\begin{enumerate}[leftmargin=*]
  \item open-ended conversation;
  \item optional Shared Symbol proposals;
  \item meaning negotiation;
  \item a period without using the symbol;
  \item reuse of the symbol in a later conversation;
  \item repair if needed;
  \item review of the Private Etymology record;
  \item final interpretation and relationship measures.
\end{enumerate}

\subsection{Measures}

Measures should cover four areas. \textbf{Interpretation performance} includes delayed interpretation accuracy and whether reuse fits the current context. \textbf{Relational experience} includes perceived shared history, dyadic uniqueness, relational continuity, mutual authorship, and shared understanding. \textbf{Trust and risk} include trust in provenance, willingness to interact again, irritation or cognitive burden, emotional responses to incorrect reuse, and perceived manipulation. \textbf{Control usability} asks whether users can understand, revise, retire, and delete symbols and their provenance without too much effort.

Qualitative analysis should examine how participants describe moments when a symbol ``felt like ours,'' when it felt forced by the model, and when an expression that was once shared no longer fit.

\subsection{Evaluation of Private Etymology}

Private Etymology should be evaluated not only as a memory mechanism but also as an accountability interface. Participants could review the current meaning, origin summary, selected evidence turns, revision sequence, provenance status, and available edit or deletion operations.

Researchers can then examine whether this representation increases trust and control, or whether it makes the interaction too formal and reduces the informal quality of personal idioms.

\section{Ethical and Design Considerations}

\subsection{Privacy and Visibility}

A dyad-specific meaning can feel private in a social sense even when it is not technically confidential. Systems should not treat these two kinds of privacy as the same.

Private Etymology may include sensitive emotional events, relationship conflicts, health information, sexual references, or other intimate material. The system should minimize storage, protect the data, and give the user control over it.

A glanceable interface must allow per-symbol visibility settings.

\subsection{Consent and Mutuality}

An AI system cannot assume mutual acceptance only because the user did not object. Silence may mean confusion, fatigue, or indifference.

A system should treat a symbol as established only when there is stronger evidence, such as explicit confirmation, user reuse, or successful interpretation at a later time.

\subsection{Transparency of Machine Participation}

A conversational AI does not need to be conscious for co-created symbolic interaction to be enjoyable or meaningful to the user. The design value comes from the interaction, interpretation, and continuity.

However, the system should not present model inference as if it were lived experience. It should distinguish:

\begin{quote}
\emph{The record shows that we used this expression before.}
\end{quote}

from:

\begin{quote}
\emph{I personally remember feeling what you felt that day.}
\end{quote}

The first statement can be supported by provenance. The second makes a claim about the system's inner experience that the system cannot support with evidence.

\subsection{Deletion, Exit, and Relational Change}

Relationships change. Users must be able to leave a relationship framing without repeatedly seeing symbols from that earlier framing.

A system should support individual-symbol retirement, bulk deletion, relationship reset, export, selective forgetting, and suspension of proactive reuse.

Past mutual agreement does not mean permanent consent.

\section{Discussion}

\subsection{Shared Memory Is Not a Shared Symbol}

A shared memory is knowledge that an earlier event happened.

A Shared Symbol is a compact expression whose use has become connected, within a dyad, to a situation, stance, action, or relational implication.

Private Etymology is the relational provenance that records how this connection formed and changed.

\begin{figure}[!ht]
\centering
\includegraphics[width=0.92\columnwidth]{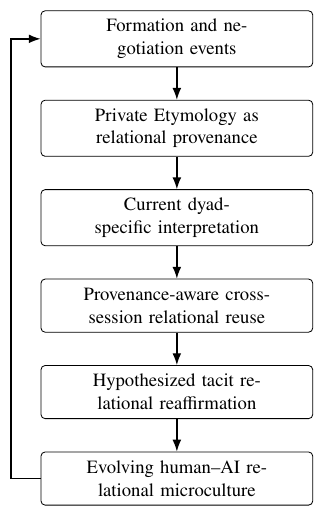}
\caption{Private Etymology records the history of a Shared Symbol and supports its reuse across conversations. Relational reaffirmation is a hypothesis, not an established result.}
\label{fig:model}
\end{figure}

Figure~\ref{fig:model} summarizes the proposed design account.

A memory can exist without becoming a symbol. A symbol can also remain in use after the details of its original memory have been forgotten. Private Etymology connects these two without treating them as the same thing.

\subsection{From Relationship Definition to Relational Microculture}

The earlier Mikasa work treated character coherence and relationship definition as basic structural responsibilities of an AI companion.

The present work extends that position.

Relationship framing can define who the participants are to each other, which interaction norms apply, and what tone or level of intimacy is appropriate.

However, it cannot define in advance the specific microculture that may later develop. That content must grow through interaction.

Private Etymology offers one way for a computational system to support this development. It allows the relationship to create expressions that were not fully specified in the initial persona or prompt, while keeping evidence of how those expressions gained meaning.

This distinction is important for relationship-centered AI design.

A completely fixed relationship cannot develop. A completely unstructured relationship leaves the user to maintain coherence alone. Relationship-first design can provide a stable frame, while Private Etymology supports new local content that emerges inside that frame.

\subsection{Theoretical Contribution}

This paper does not claim that Shared Symbols, personal idioms, relationship microcultures, symbolic co-construction, delayed reuse, or memory provenance are individually new.

Its theoretical contribution is to connect them in the following design mechanism:

\begin{center}
Formation and negotiation events $\rightarrow$ Private Etymology as relational provenance $\rightarrow$ current dyad-specific interpretation $\rightarrow$ provenance-aware cross-session relational reuse $\rightarrow$ hypothesized tacit reaffirmation of shared history.
\end{center}

The key point is to treat the form--meaning relationship itself, not only the original factual event, as a persistent, revisable, and evidence-grounded system object.

\subsection{Technical Contribution}

Private Etymology extends conversational memory in four main ways.

First, it is \textbf{symbol-centered} rather than only fact-centered.

Second, it represents \textbf{multi-layered interpretation}, including propositional, affective, pragmatic, and relational dimensions.

Third, it records \textbf{mutuality and disagreement}, distinguishing proposal, acceptance, repair, contestation, and retirement.

Fourth, it constrains \textbf{future action} by specifying whether and where the expression may be reused.

In this way, provenance can guide future system action instead of serving only as an archive.

\subsection{Limitations}

The full Private Etymology model and the relational-reaffirmation claim are still conceptual, although the conservative symbol-admission process and Watch display have been implemented.

The literature comparison in this paper is not a systematic review. Closely related work may use other terms, such as inside jokes, dyadic traditions, private language, conceptual pacts, lexical entrainment, relational culture, shared reality, or convention evolution.

Private Etymology may also be too formal for some interactions. Human personal idioms often emerge without explicit awareness, and showing their full provenance may reduce their spontaneity.

The proposed Private Etymology schema is only an example and may be more complex than necessary. The current prototype deliberately uses a much smaller snapshot. It keeps only the latest glyphs and timestamps. Meanings, older symbols, lifecycle states, provenance events, mutual-uptake counts, and Private Etymology are not yet stored. Persisted-symbol reuse is approximated by exact equality with the latest saved glyph string, and user retraction does not yet clear the current complication.

The model still classifies evidence types and Boolean factors, so classification errors are possible. The additive weights and the threshold of 80 are design heuristics, not empirically calibrated probabilities. These limitations should be tested before this policy is used beyond the current prototype.

The relational-reaffirmation hypothesis has not yet been tested. Successful recognition may increase perceived continuity, but it may also have no effect or may make users uncomfortable when they become aware of how the system stores the information.

Finally, this paper focuses on one human--AI dyad. Group microcultures, multi-agent systems, and expressions shared by several humans and AI agents need separate analysis.

\balance
\section{Conclusion}

This paper introduced Private Etymology as a provenance-aware concept for long-term human--AI relationships. It also reported an intentionally limited Apple Watch prototype for Shared Symbols.

Private Etymology is not only a record of when a symbol first appeared. It is a machine-representable relational provenance that describes how a dyad-specific symbolic expression was proposed, interpreted, negotiated, repaired, reused, revised, stabilized, contested, forgotten, or retired over time.

The ordered events in that provenance form an interactional trajectory. Provenance remains the main concept because the design problem is not only about change over time. It also concerns evidence, authorship, mutuality, uncertainty, and accountability.

The proposal builds on established research showing that humans form shared symbols, conceptual pacts, personal idioms, relationship symbols, and microcultures together. It also builds on newer work on human--AI symbolic meaning co-construction, relational agents, and long-term conversational memory.

The contribution is to turn these ideas into a system model for relational reuse. A Shared Symbol may compress propositional content, affective stance, pragmatic force, and relational implication into a compact form. When a Shared Symbol is reused in a later conversation and the other participant still understands it, this may remind both sides that the meaning comes from their shared history. The prototype provides a first safety mechanism for this process. The language model classifies the conversational evidence, but local code makes the final update decision. Only the accepted symbol data is sent to the wearable device.

This possibility needs careful handling. Systems should not invent origins, infer mutual agreement from silence, keep unwanted meanings forever, or present stored provenance as if the system personally remembered the experience. Private Etymology should remain inspectable, contestable, revisable, forgettable, and deletable.

Instead of treating relationship-specific symbols as isolated conventions or decorative personalization, this paper treats them as reusable building blocks that may support an evolving human--AI relational microculture.

\clearpage
\onecolumn
\appendix
\section{Complete Illustrative Private Etymology Schema}
\label{app:full-schema}

The following JSON is an illustrative design object, not a normative standard. It shows the difference between the current interpretation, evidence-grounded relational provenance, meaning change, and future reuse policy.

\begin{lstlisting}[style=jsonstyle,basicstyle=\ttfamily\scriptsize,aboveskip=0.4em,belowskip=0.4em]
{
  "symbol_id": "ss_001",
  "dyad_id": "human_01__agent_mikasa",
  "form": {
    "value": "[raccoon]",
    "modality": "emoji"
  },
  "status": "established",
  "current_interpretation": {
    "propositional_content":
      "The situation could not reasonably have been avoided.",
    "affective_stance":
      "Light resignation with mild amusement.",
    "pragmatic_force":
      "Accept the situation without assigning blame.",
    "relational_implication":
      "We recognize this as one of our recurring absurd situations."
  },
  "private_etymology": {
    "provenance_status": "evidence_grounded",
    "origin_summary":
      "First used after an absurd but unavoidable mishap.",
    "events": [
      {
        "event_id": "pe_001",
        "event_type": "contextual_coining",
        "actor": "ai",
        "timestamp": "YYYY-MM-DDThh:mm:ss",
        "evidence_ref": "conversation_turn_184",
        "proposed_interpretation":
          "That cannot be helped, with light humor."
      },
      {
        "event_id": "pe_002",
        "event_type": "human_reuse",
        "actor": "human",
        "timestamp": "YYYY-MM-DDThh:mm:ss",
        "evidence_ref": "conversation_turn_229",
        "interpretation_delta":
          "Extended from one mishap to a recurring category."
      },
      {
        "event_id": "pe_003",
        "event_type": "mutual_confirmation",
        "actor": "dyad",
        "timestamp": "YYYY-MM-DDThh:mm:ss",
        "evidence_ref": "conversation_turn_231"
      }
    ],
    "human_confirmed": true,
    "last_reviewed_at": "YYYY-MM-DDThh:mm:ss"
  },
  "meaning_history": [
    {
      "version": 1,
      "status": "superseded",
      "gloss": "That cannot be helped."
    },
    {
      "version": 2,
      "status": "current",
      "gloss":
        "An absurd but unavoidable situation accepted with humor."
    }
  ],
  "reuse_policy": {
    "reuse_allowed": true,
    "require_confirmation_after_days": 180,
    "allowed_surfaces": ["watch_complication", "chat"],
    "privacy_level": "private_to_user"
  },
  "last_mutually_understood_at": "YYYY-MM-DDThh:mm:ss"
}
\end{lstlisting}

\clearpage
\twocolumn
\bibliographystyle{plainnat}
\bibliography{private_etymology}

\end{document}